\documentclass[twocolumn,superscriptaddress,preprintnumbers,amsmath,amssymb,prc,10pt]{revtex4-2} 
\usepackage{graphicx}% Include figure files
\usepackage{dcolumn}% Align table columns on decimal point
\usepackage{bm}% bold math
\usepackage{float}
\usepackage{color}
\usepackage{CJK}% Chinese character support
\usepackage[colorlinks,linkcolor=blue,urlcolor=blue,citecolor=blue]{hyperref}
\usepackage{isotope}
\usepackage{amsmath}
\usepackage[normalem]{ulem}
\renewcommand{\sout}{\bgroup \color{red} \ULdepth=-0.5ex \ULset}

\usepackage{scalerel} 
\allowdisplaybreaks[4]

\makeatletter
\renewcommand{\maketag@@@}[1]{\hbox{\m@th\normalsize\normalfont#1}}%
\makeatother
\usepackage{tikz,xcolor}
\definecolor{lime}{HTML}{A6CE39}
\DeclareRobustCommand{\orcidicon}{
	\begin{tikzpicture}
	\draw[lime, fill=lime] (0,0) 
	circle [radius=0.16] 
	node[white] {{\fontfamily{qag}\selectfont \tiny ID}};
	\draw[white, fill=white] (-0.0625,0.095) 
	circle [radius=0.007];
	\end{tikzpicture}
	\hspace{-2mm}}

\foreach \x in {A, ..., Z}{%
	\expandafter\xdef\csname orcid\x\endcsname{\noexpand\href{https://orcid.org/\csname orcidauthor\x\endcsname}{\noexpand\orcidicon}}
	}

\begin{document}
\begin{CJK*}{UTF8}{gbsn}% Use default fonts from CJK (see below)
%\linenumbers
%\preprint{APS/123-QED}

\title{Constraining neutron-proton effective mass splitting through nuclear giant dipole resonance within transport approach}

\author{Yi-Dan Song(宋一丹)\orcidA{}}
\email{songyidan@usst.edu.cn}
\affiliation{College of Science, University of Shanghai for Science and Technology, Shanghai $200093$, China}

\author{Min-Si Luo(罗闽斯)\orcidE{}}
\affiliation{College of Science, University of Shanghai for Science and Technology, Shanghai $200093$, China}

\author{Rui Wang({王睿})\orcidB{}}
%\thanks{Present address: INFN, Laboratori Nazionali del Sud, I-$95123$ Catania, Italy}
\email{rui.wang@lns.infn.it}
\affiliation{Istituto Nazionale di Fisica Nucleare (INFN), Laboratori Nazionali del Sud, I-$95123$ Catania, Italy}
\affiliation{INFN, Sezione di Catania, I-95123 Catania, Italy}
\affiliation{Shanghai Research Center for Theoretical Nuclear Physics, NSFC and Fudan University, Shanghai $200438$, China}

\author{\\Zhen Zhang(张振)\orcidC{}}
\email{zhangzh275@mail.sysu.edu.cn}
\affiliation{Sino-French Institute of Nuclear Engineering and Technology, Sun Yat-sen University, Zhuhai $519082$, China}
\affiliation{Shanghai Research Center for Theoretical Nuclear Physics, NSFC and Fudan University, Shanghai $200438$, China}

\author{Yu-Gang Ma(马余刚)\orcidD{}}
\email{mayugang@fudan.edu.cn}
\affiliation{Key Laboratory of Nuclear Physics and Ion-beam Application~(MOE), and Institute of Modern Physics, Fudan University, Shanghai $200433$, China}
\affiliation{Shanghai Research Center for Theoretical Nuclear Physics, NSFC and Fudan University, Shanghai $200438$, China}

\date{\today}% It is always \today, today,
\begin{abstract}
{Based on the Boltzmann-Uehling-Uhlenbeck equation, we investigate the effects of the isovector nucleon effective mass $m^*_{v,0}$ and the in-medium nucleon-nucleon cross section $\sigma^*$ on the isovector giant dipole resonance~(IVGDR) in $\isotope[208]{Pb}$, employing a set of representative Skyrme energy density functionals. 
We find that the energy-weighted sum rule $m_1$ of the IVGDR is highly sensitive to $m^{*}_{v,0}$ and only mildly dependent on $\sigma^*$, while the width $\Gamma$ of the IVGDR is primarily governed by $\sigma^*$ with a moderate sensitivity to $m^*_{v,0}$. 
From a Bayesian analysis of both $m_1$ and $\Gamma$, we infer the isovector effective mass $m^{*}_{v,0}/m$ = $0.731^{+0.027}_{-0.023}$, where $m$ is the bare nucleon mass. 
Furthermore, by incorporating the isoscalar effective mass $m^*_{s,0}/m = 0.820 \pm 0.030$, extracted from the isoscalar giant quadrupole resonance in $\isotope[208]{Pb}$, the linear neutron-proton effective mass splitting coefficient at saturation density $\rho_0$ is determined to be $\Delta m^*_1 (\rho_0)/m = 0.200 ^{+0.101}_{-0.094}$.}
\end{abstract}

\maketitle
\section{Introduction}\label{1}
The concept of nucleon effective mass $m^{*}$, first introduced by Brueckner, characterizes the momentum and/or energy dependence of the single-nucleon potential in nuclear matter~\cite{BruPR97, JeuPR25, LBAPPNP99}.
It allows one to describe the motion of a nucleon with bare mass $m$ in the momentum-dependent potential as equivalent to the motion of a nucleon with effective mass $m^{*}$ in a momentum-independent potential.
In isospin-asymmetric nuclear matter, the difference between neutron ($n$) and proton ($p$) potentials may lead to a neutron-proton effective mass splitting $m^*_{n-p}$ $\equiv$ $m^*_n-m^*_p$~\cite{MeiEPJA31, MeiEPJA32}.
The $m^*_{n-p}$ plays an important role in understanding various phenomena, including the transport properties of asymmetric nuclear matter~\cite{XCPRC82,LBAMPLA30,XJPRC91}, the dynamical evolution of heavy-ion collisions~(HICs)~\cite{LBAPR464,FZQNPA878,FZQNST24,GYFCPC41,ZFNST31,WFYNST34,ZYXPLB732,YJPRC109,YJaXv2506,MorPLB799,TLCPC44,TsaPLB853}, neutrino emission in neutron stars~\cite{BalPRC89}, and the primordial nucleosynthesis in the early universe~\cite{SteIJMPE15}.
Accurate constraints on $m^*_{n-p}$ are thus essential for a deeper understanding of the above topics. 

Over the past few decades, substantial efforts have been made to determine $m^*_{n-p}$ through various $ab$ $initio$ or microscopic approaches~\cite{BalPRC89,WSBPRC108},  as well as through phenomenological extractions based on experimental observables~\cite{LBAMPLA30,LBAPLB727}.
However, both quantitative and qualitative consensus on the behavior of $m^*_{n-p}$ in isospin-asymmetric nuclear matter remains elusive.
In particular, analyses of various observables from nuclear structure and reaction experiments have led to inconsistent constraints on $m^*_{n-p}$.
For example, constraints on the first-order~(linear) neutron-proton effective mass splitting coefficient $\Delta m^*_1(\rho_0)$ $\equiv$ $\frac{\partial m^*_{n-p}(\rho_0)}{\partial \delta}\Big|_{\delta=0}$ at nuclear saturation density $\rho_0$ span a wide range from $-0.13m$ to $0.56m$, demonstrating a strong dependence on the specific model and experimental probe~\cite{MorPLB799, TLCPC44, TsaPLB853, LBAPLB727, LXHPLB743, ZZPRC93, KHYPRC95, XJPRC102,ZZCPC45,WSBPRC108}.
Among these results, studies based on nucleon-nucleus scattering and nuclear collective motions consistently suggest a positive $\Delta m^{*}_{1}(\rho_{0})$~\cite{LBAPLB727, LXHPLB743, ZZPRC93, KHYPRC95, XJPRC102, ZZCPC45}, while analyses based on HICs yield conflicting conclusions~\cite{MorPLB799, TLCPC44, TsaPLB853, ZYXPLB732, YJPRC109, YJaXv2506}.
Similarly, theoretical predictions from \textit{ab initio} calculations show significant variation across different methods and interactions~\cite{WSBPRC108}.
Under these circumstances, any new and independent constraints on $m^*_{n-p}$ would be valuable for clarifying the debate and advancing toward a definitive conclusion about its behavior in isospin-asymmetric nuclear matter.

The nuclear giant resonances, especially the isovector giant dipole resonance~(IVGDR) which can either be photon excited \cite{Lan,Sun,Jiao} or heavy-ion excited as well as the isoscalar giant quadrupole resonance~(ISGQR), are highly sensitive to nucleon effective masses~\cite{ZZPRC93,ZZCPC45,KHYPRC95,XJPLB810,XJPRC102,LZZPRL131,SYDPRC104}. 
The ISGQR primarily provides information on the isoscalar nucleon effective mass $m^{*}_{s}$, i.e., the nucleon effective mass in symmetric nuclear matter.
On the other hand, the IVGDR is commonly used to probe the isovector nucleon effective mass $m^{*}_{v}$, corresponding to the neutron~(proton) effective mass in pure proton~(neutron) matter.
Constraints on $m^{*}_{s}$ and $m^{*}_{v}$ from the ISGQR and IVGDR, respectively, then enable the extraction of the $\Delta m^*_1$ and the $m_{n-p}^*$.
Currently, various methods are employed to calculate the observables of nuclear giant resonances, including the (quasi-)RPA approach~\cite{ZZPRC93, ZZCPC45, XJPLB810, LZZPRL131,Colo}, and dynamical approaches such as the Boltzmann-Uehling-Uhlenbeck~(BUU) equation~\cite{KHYPRC95,XJPRC102,SYDPRC104}, the quantum molecular dynamic model~\cite{HWBPRL113,HBSPRC103}, and the macroscopic Lagnevin equation~\cite{SJPRC101,WXPRC104}.
Among these models, the BUU transport equation, which describes the time evolution of one-body phase-space distribution functions~\cite{BerPR160,BonPR243,BusPR512,XJPRC93,ZYXPRC97,OnoPRC100,ColPRC104,WolPPNP125}, is widely used in studies of HICs and nuclear giant resonances.
It has succeeded in describing various experimental observables such as particle spectra in nucleus collisions and strength functions of nuclear collective motions, and thus serves as an effective tool in constraining $m^*_{n-p}$.
 
In the present work,the BUU equation is employed to investigate systematically the effects of the $m^*_{v}$ and the in-medium nucleon-nucleon~(NN) cross sections $\sigma^*$ on the IVGDR in $\isotope[208]{Pb}$.
The energy-weighted sum rule $m_1$ and the width $\Gamma$ of IVGDR are calculated by implementing a set of representative Skyrme energy density functionals~(EDF).
Based on these results, we further employ the Bayesian inference to extract the value of $m^*_{v,0}$, i.e., the isovector nucleon effective mass at $\rho_0$.
The resulting $m^*_{n-p}$, deduced by combining existing constraints on $m^*_{s,0}$~(the isoscalar nucleon effective mass at $\rho_0$), is then compared with the values obtained from other phenomenological analyses and {\it ab initio} calculations.
In Sec.~\ref{2}, we introduce  the theoretical framework used in this work, including the treatment of nuclear collective motions using the BUU equation and nucleon effective masses within the Skyrme EDF.
The results and their corresponding discussions are presented in Sec.~\ref{3}, followed by a summary in Sec.~\ref{4}.

\section{Theory and Method}\label{2}

\subsection{Boltzmann-Uehling-Uhlenbeck equation and nuclear collective motions}

The BUU equation describes the time evolution of the Wigner function $f({\bf r},{\bf p})$, which can be obtained by applying a semiclassical approximation to time-dependent quantum approaches~\cite{DanAP152,BerPR160,BonPRL71,BusPR512}.
For describing nuclear collective motions, only the Wigner functions for neutrons and protons are need, and their BUU equations read,
\begin{equation}\label{E:BUU}
 (\partial_t +{\boldsymbol\nabla}_p\epsilon_\tau\cdot{\boldsymbol\nabla}_r -{\boldsymbol\nabla}_r\epsilon_\tau\cdot{\boldsymbol\nabla}_p)f_\tau = I_\tau^{\rm coll},
\end{equation}
where $\tau$ $=$ $n$ or $p$.
The left-hand side of the above equation describes the time evolution of $f_\tau({\bf r},{\bf p})$ under the influence of nuclear mean-field potential, where $\epsilon_\tau[{\bf r},{\bf p},f_n,f_p]$ denotes the single-particle energy.
In the present work, it is derived based on the standard Skyrme-Hartree-Fock approach~\cite{VauPRC5,ChaNPA627,ChaNPA635}.
The collision integral on the right-hand side of Eq.~(\ref{E:BUU}) consists of a gain term~($K^<_\tau$) and a loss term~($K^>_\tau$),
\begin{widetext}
\begin{equation}\label{E:Ic}
\begin{split}
I^{\rm coll}_\tau[f_n,f_p] &= K_\tau^{<}[f_n,f_p](1-f_\tau) - K_\tau^{>}[f_n,f_p]f_\tau\\
& = g\int\prod_{i}^{2,3,4}\frac{{\rm d}{\bf p}_i}{(2\pi\hbar)^{3}}|\mathcal{M}_{12\leftrightarrow34}|^{2}(2\pi)^{4}\delta^{4}(p+p_2-p_3-p_4)\big[f_{3}f_{4}(1-f_\tau)(1-f_{2})-f_\tau f_2(1-f_3)(1-f_4)\big],
\end{split}
\end{equation}
\end{widetext}
where $f_i$ with $i$ $=$ $2,3,4$ in the second line is short for $f_{\tau_i}({\bf r},{\bf p}_i)$, and the ($1 - f_i$) accounts for the Pauli principle of nucleons.
The $g$ $=$ 2 is the spin degeneracy factor, and $\mathcal{M}_{12\leftrightarrow34}$ is the in-medium transition matrix element of two-body elastic NN scatterings.
The $|\mathcal{M}_{12\leftrightarrow34}|^2$ is deduced from the in-medium NN scattering cross section $\sigma^{*}$, which is approximated by the free-space NN scattering cross section $\sigma^{\rm free}$~\cite{CugNIMB111} multiplied by a constant in-medium correction factor $k$~\cite{LopPRC90}.

In this work, we solve the BUU equation utilizing the test particle ansatz, i.e., mimicking $f_\tau({\bf r},{\bf p},t)$ with a large number of test particles~\cite{WonPRC25},
 \begin{equation}\label{E:f_TP}
f_\tau({\bf r},{\bf p},t) = \frac{1}{g}\frac{(2\pi\hbar)^{3}}{N_{E}}\sum^{AN_{E}}_{i\in \tau}S[{\bf r}_{i}(t)-{\bf r}]\delta[{\bf p}_{i}(t)-{\bf p}].
\end{equation}
In the above equation, ${\bf r}_i(t)$ and ${\bf p}_i(t)$ denote the coordinate and momentum of test particle $i$, respectively.
$A$ is the mass number of the system, and $N_E$ is the number of parallel ensembles~(or the number of test particles in some literature).
The summation runs over all test particles with isospin $\tau$.
To reduce numerical fluctuations in the calculation of single-particle energies $\epsilon_\tau$, a triangular profile function $S$ for test particles is introduced in coordinate space.
We employ the lattice Hamiltonian method~\cite{LenPRC39,WRPRC99} to deal with the mean-field evolution and the stochastic collision method~\cite{DanNPA533,XZPRC75} to handle the collision integral.
A ground-state nucleus at zero temperature can be described by the static solution of Eq.~(\ref{E:BUU}).
The momentum distribution is described by zero-temperature Fermi distribution, this solution leads to the radial density distribution $\rho_{\tau}(r)$ of the nucleus, which can be obtained using the Thomas-Fermi approach via the variation of the total energy with respect to $\rho_{\tau}(r)$~\cite{GaiPRC81,BusPR512,WRPRC99}.
The present framework of solving the BUU equation has been successfully applied to study nuclear collective motions in heavy nuclei~\cite{WRPRC99, WRPLB807, SYDPRC104, SYDPRC108}, as well as light-nuclei yields~\cite{WRPRC108, WRaXv2507} and proton collective flows~\cite{WSPPRC111} in heavy-ion collisions.
One can find more detailed descriptions of the present framework of solving the BUU equation in Ref.~\cite{WRFiP8}.

The collective motion of a nucleus can be generated by applying a perturbation excitation operator $\hat{Q}$ to its ground state at the initial time $t_{0}$, i.e., $\hat{H}(t)$ = $\hat{H}_{0}$ + $\lambda\hat{Q}\delta(t-t_{0})$, where $\lambda$ is a small excitation parameter.
To initiate the IVGDR, the excitation operator ${{\hat{Q}}_{\scaleto{\rm IVD}{3.5pt}}}$ can be written as the sum of single-particle operators $\hat{q}_i$, i.e.,
\begin{equation}\label{E:GDR}
\hat{Q}_{\scaleto{\rm IVD}{3.5pt}} = \sum_i^A\hat{q}_i,\quad{\rm with}~~\hat{q}_i = \begin{cases}
    \frac{Z}{A}\hat{z}_{i},\quad {\rm for~neutrons},\\
    \frac{N}{A}\hat{z}_{i},\quad {\rm for~protons},
    \end{cases}
\end{equation}
where $\hat{z}_{i}$ is $z$-component of the coordinate operator of the $i$th nucleon.
The $N$ and $Z$ denote the neutron and proton numbers in the nucleus, respectively.

In the Wigner representation, the expectation value of excitation operator $\langle\hat{Q}_{\scaleto{\rm IVD}{3.5pt}}\rangle$ can be expressed as
\begin{equation}\label{E:Q2}
\langle\hat{Q}_{\scaleto{\rm IVD}{3.5pt}}\rangle = \int\sum_{\tau=n,p}{f_\tau(\vec{r},\vec{p})q_\tau(\vec{r},\vec{p})d^{3}\vec{r}d^{3}\vec{p}},
\end{equation}
where the $q_\tau(\vec{r},\vec{p})$ is Wigner transform of $\hat{q}_i$ for neutrons or protons.
The detailed derivation of relevant equations can be found in Ref.~\cite{UrbPRC85,WRPRC99,WRPLB807,SYDPRC104,SYDPRC108}.

Within the framework of the BUU equation, the isovector dipole excitations can be generated by perturbing the initial momenta ${\bf p}_i$ of test particles with respect to their values in the ground state,
\begin{equation}\label{E:ex}
p_{z,i}\rightarrow
\begin{cases}
~{p_{z,i}}-{\lambda}\frac{N}{A},\quad{\rm for~protons},\\
~{p_{z,i}}+{\lambda}\frac{Z}{A},\quad{\rm for~neutrons}.
\end{cases}
\end{equation}
Following the excitation of the nucleus as described in the above equation, we can obtain the time evolution of the expectation $\langle\hat{Q}\rangle(t)$ for the IVGDR, and subsequently calculate the strength function $S(E)$ through the Fourier transform of $\Delta\langle\hat{Q}\rangle(t)$ within the framework of line response theory~\cite{Fet1971},
\begin{equation}\label{E:SE}
S(E) = -\frac{1}{\pi\lambda}{\int^{\infty}_{0}}{dt\Delta\langle{\hat{Q}}\rangle(t){\rm sin}\frac{Et}{\hbar}}.
\end{equation}
The $\Delta\langle\hat{Q}\rangle(t)$ $=$ $\langle0^{'}|\hat{Q}|0^{'}\rangle$ $-$ $\langle0|\hat{Q}|0\rangle$ denotes the time evolution of the response function of the nucleus to the $\hat{Q}$.
The $|0\rangle$ and $|0^{'}\rangle$ are the  states before and after the perturbation, respectively.

Several quantities are defined based on the strength function $S(E)$, including its full width at half maximum~(FWHM), or simply the width $\Gamma$, as well as its energy moments,
\begin{equation}\label{E:mk}
m_n = \int{E^{n}S(E)dE}.
\end{equation}
In particular, when $n = 1$, the above equation yields the first moment $m_1$, commonly referred to as the energy-weighted sum rule.

\subsection{Nucleon effective masses and neutron-proton effective mass splitting}

Within the standard Skyrme EDF~\cite{VauPRC5, ChaNPA627, ChaNPA635}, the nucleon effective mass $m^*_\tau$ for isospin state $\tau$~(either $n$ or $p$) can be expressed in terms of the neutron density $\rho_n$ and proton density $\rho_p$,
\begin{equation}\label{E:m}
\begin{split}
\frac{\hbar^2}{2m^*_{\tau}(\rho,\delta)}=&~\frac{\hbar^2}{2m}+\frac{1}{4}t_{1}\Big[(1+\frac{1}{2}x_{1})\rho-(\frac{1}{2}+x_{1})\rho_\tau\Big]\\
&+\frac{1}{4}t_{2}\Big[(1+\frac{1}{2}x_{2})\rho+(\frac{1}{2}+x_{2})\rho_\tau\Big],
\end{split}     		
\end{equation}
{where $\rho$ $=$ $\rho_n$ $+$ $\rho_p$, and} $x_{1}$, $t_{1}$, $x_{2}$, and $t_{2}$ are Skyrme parameters. 
By setting $\rho_\tau$ $=$ $\rho/2$, we obtain the isoscalar nucleon effective mass, i.e., the nucleon effective mass in symmetric nuclear matter,
\begin{equation}\label{E:ms}
\frac{\hbar^2}{2m^*_s(\rho)} = \frac{\hbar^2}{2m}+\frac{3}{16}t_{1}\rho+\frac{1}{16}t_{2}(5+4x_{2})\rho.
\end{equation}
The isovector nucleon effective mass, corresponding to the proton~(neutron) effective mass in pure neutron~(proton) matter, can be obtained by setting $\rho_\tau$ $=$ $0$, 
\begin{equation}\label{E:mv}
\frac{\hbar^2}{2m^*_v(\rho)} = \frac{\hbar^2}{2m}+\frac{1}{8}t_{1}(2+x_{1})\rho+\frac{1}{8}t_{2}(2+x_{2})\rho.
\end{equation}

The neutron-proton effective mass splitting can be obtained through Eqs.~(\ref{E:ms}) and (\ref{E:mv}), i.e.,  
%\begin{small}
\begin{eqnarray}\label{E:mn}
\begin{split}
m^*_{n-p}(\rho,\delta) &\equiv m^*_{n}-m^*_{p}=2m^*_{s}\sum^{\infty}_{n=1}(\frac{m^*_{s}-m^*_{v}}{m^*_{v}}\delta)^{2n-1}\\
&= \sum^{\infty}_{n=1}\Delta{m^*_{2n-1}}(\rho)\delta^{2n-1},
\end{split}  
\end{eqnarray}
%\end{small}
where $\delta$ $=$ $(\rho_{n}-\rho_{p})/(\rho_{n}+\rho_{p})$ denote the isospin asymmetry.
Specifically, the first-order~(linear) neutron-proton effective mass splitting coefficient is expressed as,
\begin{equation}\label{E:mn1}
\Delta{m^*_{1}}(\rho)\equiv\frac{\partial m^*_{n-p}(\rho)}{\partial \delta}\bigg|_{\delta=0}=2m^*_{s}(\frac{m^*_{s}-m^*_{v}}{m^*_{v}}).
\end{equation}
The above expressions indicate that within the Skyrme EDF framework, the $m^*_{n-p}$ is directly  related to $m^*_{s}$ and $m^*_{v}$. 
In isospin-asymmetric nuclear matter, the sign of $m^*_{n-p}$ corresponds to that of the difference $m^*_{s}$ $-$ $m^*_{v}$.
%In the following, we use $m^*_{s,0}$ and $m^*_{v,0}$ as a shorthand for $m^*_{s}(\rho_0)$ and $m^*_{v}(\rho_0)$, respectively.

The energy-weighted sum rule $m_1$ of the IVGDR is correlated with the isovector effective mass $m^{*}_{ v,0}/m$~\cite{ZZPRC93,ChaNPA627}, i.e.,
\begin{equation}\label{E:m1}
m_1 \approx \frac{9}{4\pi}\frac{\hbar^2}{2m} \frac{NZ}{A}( \frac{m}{m^*_{v,0}}).
\end{equation}
By comparing observables of IVGDR calculated using the BUU equation with experimental data~\cite{DieADNDT38, RocPRC92, TamPRL107}, we can constrain the $m^*_{v,0}$, thereby gaining insight into the characteristics of neutron-proton effective mass splitting.

\begin{figure*}[htb]
\centering
\includegraphics%[width=140mm]
[width=\linewidth]{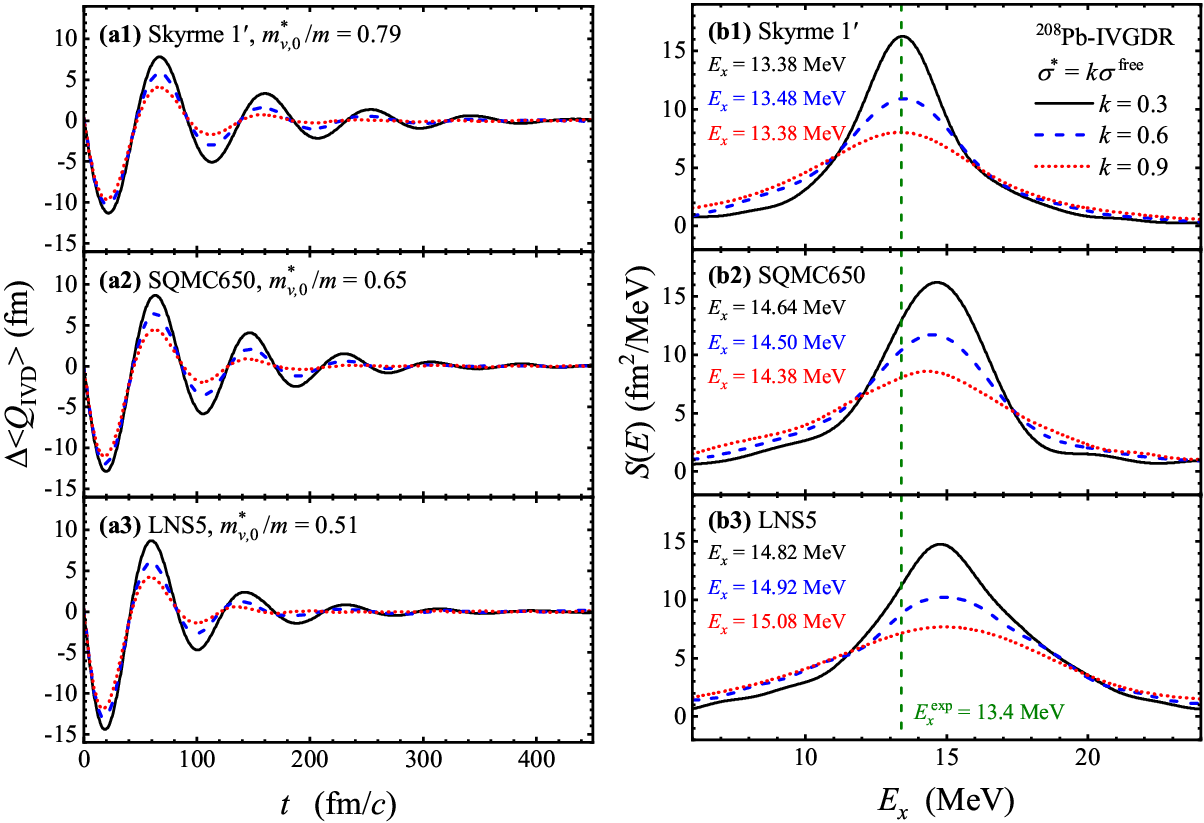}
\caption{The time evolution of the $\Delta\langle\hat{Q}_{\scaleto{\rm IVD}{3.5pt}}\rangle$ (left) and the strength function $S(E)$ (right) for IVGDR in $\isotope[208]{Pb}$ calculated based on the BUU equation with three Skyrme interactions (from top to bottom).
The black solid line, blue dashed line, and red dotted line represent the results employing different in-medium NN cross section $\sigma^*$ $=$ $k\sigma^{\rm free}$ with $k$=0.3, $k$=0.6, and $k$=0.9, respectively.
The vertical green dashed line represents the experimental peak energy $E_{x}$ $=$ $13.4~\rm MeV$ measured at RCNP~\cite{TamPRL107}.
} \label{F:Q1}
\end{figure*}

\section{Results and discussion}\label{3}

In the present study, we solve the BUU equation to obtain the strength function $S(E)$ of the IVGDR of $\isotope[208]{Pb}$.
In order to guarantee the convergence of the obtained results, we adopt a large number of ensembles with $N_{E}$ $=$ $10000$.
The perturbation excitation parameter $\lambda$ is set to be $0.015~{\rm GeV}/c$, which is small enough to ensure the applicability of the linear response theory, yet large enough to suppress numerical fluctuations.

\subsection{Sensitivity of the energy-weighted sum rule and width of IVGDR to isovector nucleon effective mass}

In order to illustrate the characteristics of the IVGDR obtained by solving the BUU equation, we first present examples of the response function $\Delta\langle{\hat{Q}}_{\scaleto{\rm IVD}{3.5pt}}\rangle$ and the strength function $S(E)$ obtained using three Skyrme interactions with distinct $m^*_{v,0}$.
In the left window of Fig.~\ref{F:Q1}, we show the time evolution of the $\Delta\langle{\hat{Q}}_{\scaleto{\rm IVD}{3.5pt}}\rangle$ for $\isotope[208]{Pb}$, obtained by solving the BUU equation employing three Skyrme interactions, namely, Skyrme $1'$, SQMC$650$, and LNS$5$ (information of these Skyrmes interactions can be found in Ref.~\cite{DutPRC85,PetNPA584,GamPRC84}), corresponding to $m^{*}_{v,0}/m$ $=$ $0.79$, $0.65$, and $0.51$, respectively.
The black solid, blue dashed, and red dotted lines represent the results obtained using constant in-medium correction factors of $k$ $=$ $0.3$, $0.6$, and $0.9$, respectively.
It is seen from the figure that all the obtained $\Delta\langle{\hat{Q}}_{\scaleto{\rm IVD}{3.5pt}}\rangle(t)$ exhibit damped oscillations and eventually diminish to their ground-state value of zero.
In particular, since two-body dissipation caused by NN scatterings plays a prominent role in the damping of IVGDR,  the oscillation amplitude decreases more rapidly in the case with larger $k$.

In the right window of Fig.~\ref{F:Q1}, we present the corresponding strength function $S(E)$, obtained via the Fourier transform of the $\Delta\langle\hat{Q}_{\scaleto{\rm IVD}{3.5pt}}\rangle(t)$ shown in the left window.
The peak energies $E_{x}$ of the strength functions for the three Skyrme interactions and various $k$ are provided in the figure.
It is evident that $E_{x}$ increases as $m^{*}_{v,0}/m$ decreases, while remaining relatively insensitive to the value of $k$.
In particular, among the three Skyrme interactions, the $E_{x}$ associated with Skyrme $1'$~($m^{*}_{v,0}/m$ $=$ $0.79$) is in the closest agreement with the experimental value $E_{x}$ $=$ $13.4~\rm MeV$~(vertical green dashed line) measured at RCNP~\cite{TamPRL107}.
Similar dependence of the $E_{x}$ of ISGQR on the isoscalar effective mass $m^{*}_{s,0}/m$ and $\sigma^*$ have been reported in Ref.~\cite{SYDPRC104}.

\begin{figure}[htbp]
%\centering
\includegraphics%[width=\columnwidth]
[width=\linewidth]{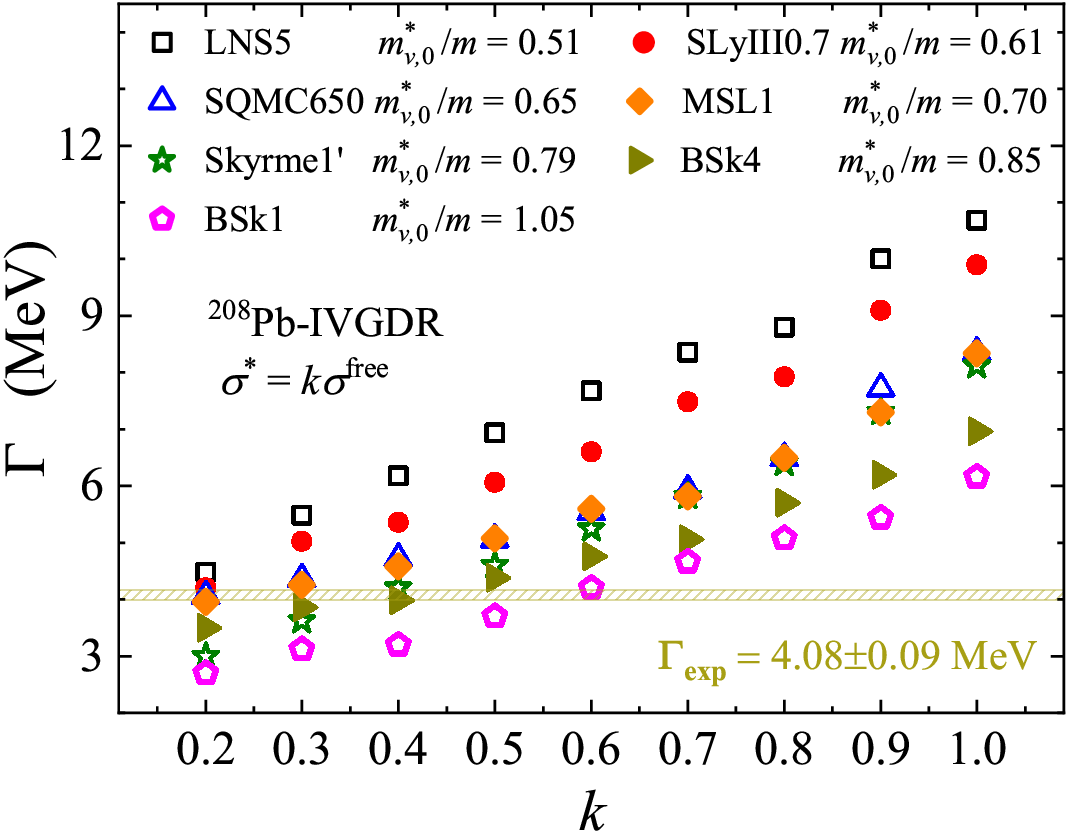}
\caption{
The $\Gamma$ of IVGDR in $\isotope[208]{Pb}$ calculated using the BUU equation with different constant in-medium correction factors $k$.
The various symbols correspond to $\Gamma$ values for different $m^{*}_{v,0}/m$, and the brown hatched band represents the experimental value of $\Gamma$ $=$ $4.08\pm0.09$ MeV~\cite{PluADNDT123}.
} \label{F:Q2}
\end{figure}

In contrast to its effect on $E_{x}$, increasing the in-medium correction factor $k$ leads to a pronounced increase in the FWHM $\Gamma$ of $S(E)$.
To systemically present the impact of $k$ on $\Gamma$, we show in Fig.~\ref{F:Q2} the $\Gamma$ as functions of $k$ obtained from the BUU equation employing various Skyrme interactions, with $m^{*}_{v,0}/m$ ranging from $0.51$ to $1.05$~\cite{DutPRC85,PetNPA584,GamPRC84,WasPRC86,ZZPLB726}.
The constant in-medium correction factor $k$ is varied from $0.2$ to $1.0$, representing a transition from a strongly suppressed in-medium NN cross section to the free-space cross section.
The brown hatched band denotes the experimental $\Gamma$ of $\isotope[208]{Pb}$, extracted from the total photoneutron cross section $\sigma(\gamma, sn)$~\cite{PluADNDT123}.
We note from the figure, focusing on the results of a particular Skyrme interaction, the positive dependence of the $\Gamma$ of $S(E)$ on the value of $k$, reflecting the rapidity of the damping exhibited in the left window of Fig.~\ref{F:Q1}, is clearly observed.
Furthermore, for a given value of $k$, Skyrme interactions with smaller $m^*_{v,0}/m$ tend to result in a larger $\Gamma$.
This inverse dependence of $\Gamma$ on $m^{*}_{v,0}/m$ is relatively weak at small values of $k$, but becomes more pronounced as $k$ increases.

We now turn to the sensitivities of the $m_1$ to $m^*_{v,0}$.
We calculate the $S(E)$ of IVGDR in $\isotope[208]{Pb}$ using the BUU equation employing $55$ representative Skyrme interactions.
These selected interactions~(BSk1, BSk4, BSk5, BSk7, BSk8, BSk9, BSk11, BSk14, BSk15, BSk17, FPLyon, KDE0v, LNS5, MSk5*, MSk7, MSL1, RATP, SIII, SK272, SKa, Ska35s15, Ska35s25, Ska45s20, SkM, SkM*, SkS1, SkS3, SkS4, SkSC14, SkT1, SkT2, SkT3, SkT8, SkT9, SKXce, SKXm, Sk$\chi$m*, Skxs15, Skxs20, Skyrme $1'$, SLy0, SLy1, SLy2, SLy4, SLy5, SLy7, SLy10, SLyIII.0.7, SQMC$650$, SQMC700, SV-bas, SV-K226, SV-kap02, SV-min, T, details of these interactions can be found in Refs.~\cite{DutPRC85, GamPRC84, ZZPLB726, PetNPA584, WasPRC86, ZZPLB777}) can reproduce relatively well the empirical values of the characteristic quantities of the nuclear equation of state~\cite{RocPPNP101}, while with the value of $m^{*}_{v,0}/m$ ranging from $0.51$ to $1.05$.
The $m_{1}$ of IVGDR is then obtained through Eq.~(\ref{E:mk}).

\begin{figure}[htbp]
%\centering
\includegraphics%[width=\columnwidth]
[width=\linewidth]{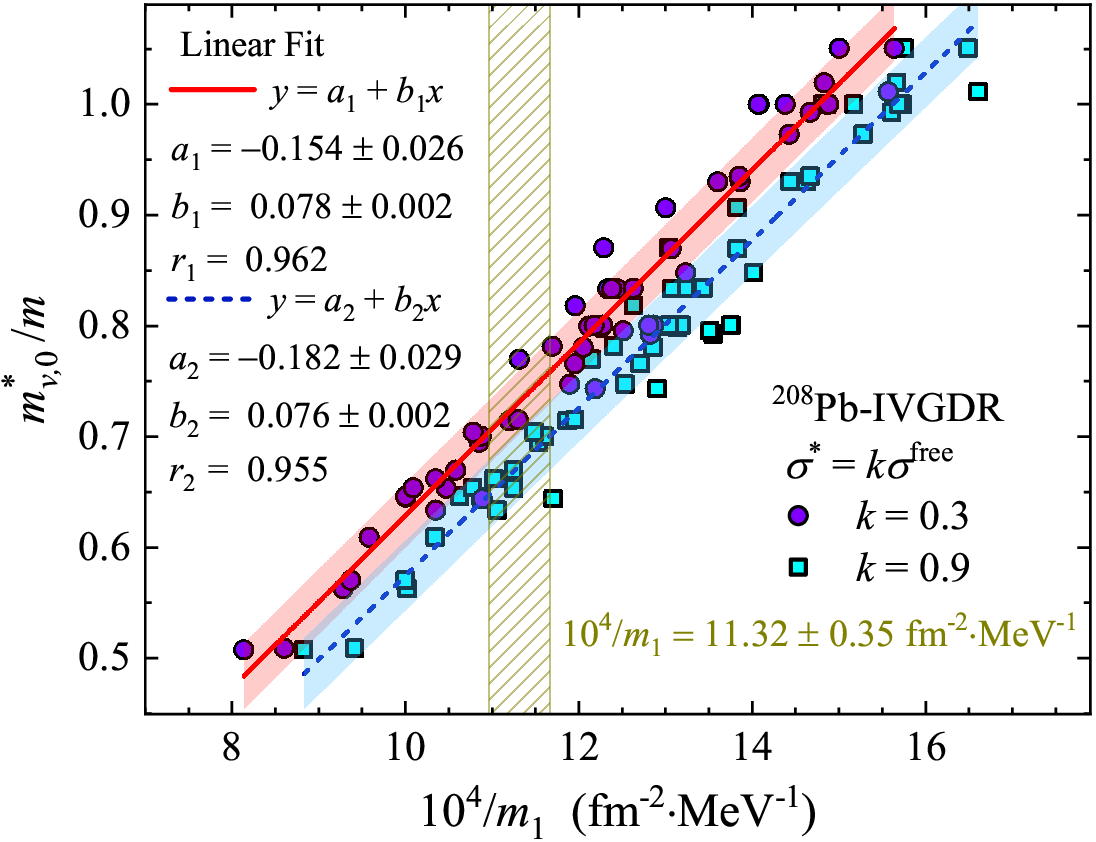}
\caption{
The correlation between $m^{*}_{v,0}/m$ and $10^{4}/m_{1}$ of IVGDR in $\isotope[208]{Pb}$. 
The purple circles and blue squares show results for $k$ $=$ 0.3 and $k$ $=$ 0.9, respectively. 
The red and blue shaded regions indicate the $68.3\%$ prediction bands of the linear regression. 
The brown hatched band represent the experimental value of $10^{4}/m_{1}$~\cite{DieADNDT38,RocPRC92,TamPRL107}.
} \label{F:Q3}
\end{figure}

Figure~\ref{F:Q3} provides the scatter plot of the $m^{*}_{v,0}/m$ and $10^{4}/m_{1}$ obtained using the BUU equation employing the above Skyrme interactions.
The purple circles and blue squares represent the results calculated adopting two different $\sigma^*$, namely, $k$ $=$ $0.3$ and $0.9$, respectively.
A clear linear correlation between $m^{*}_{v,0}/m$ and $10^{4}/m_{1}$, as indicated in Eq.~(\ref{E:m1}), is observed in the figure.
The Pearson coefficients of the correlation in this two $k$ value cases are as high as $0.962$ and $0.955$, respectively, highlighting the strong correlation between $m^{*}_{v,0}/m$ and $m_1$.
The linear fits to the results with $k=0.3$ and $k=0.9$ are shown by the red solid and blue dashed lines, respectively, with the shaded regions indicating the corresponding $1\sigma$ uncertainties.
The fluctuations of the scatterings plot with respect to the linear fits reflect mild sensitivities of $m_1$ to other quantities such as nuclear equation of states associated with those Skyrme interactions.
The brown hatched band represents the value $10^{4}/m_{1}$ $=$ 11.32 $\pm$ 0.35 $\rm fm^{-2} \cdot MeV^{-1}$, deduced from the experimental centroid energy $E_{-1}$ $\equiv$ $\sqrt{m_{1}/m_{-1}}$ and inverse energy weighted sum rule $m_{-1}$.
Here, the $E_{-1}$ $=$ 13.46 MeV is extracted from the photo-absorption measurements~\cite{DieADNDT38}, while the $m_{-1}$ is obtained from the corrected value of the electric dipole polarizability $\alpha_{D}$ $=$ 19.6 $\pm$ 0.6 $\rm fm^3$~\cite{RocPRC92, TamPRL107}.
We conclude from the figure that, despite the mild influence from the in-medium correction factor $k$, $m^{*}_{v,0}$ can still be relatively well constrained by the the energy-weighted sum rule $m_1$.

\subsection{Bayesian inference of isovector effective mass}
                                              
\begin{figure}[htbp]
%\centering
\includegraphics%[width=\columnwidth]
[width=\linewidth]{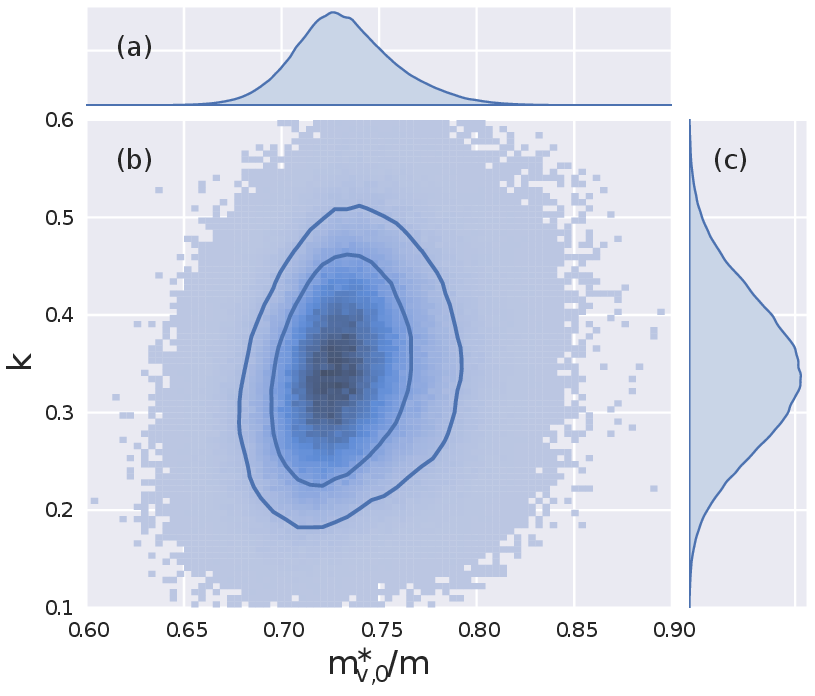}
\caption{Posterior univariate~[(a) and (c)] and bivariate~(b)
distributions of the constant in-medium correction and $m^{*}_{v,0}/m$. 
The shaded regions in panel (b) represent the 68.3\% and 90\% confidence intervals.
} \label{F:Q4}
\end{figure}

Given the sensitivity of the IVGDR width $\Gamma$ and $m_1$ to both $m_{v,0}^*$ and $k$, we further perform a Bayesian analysis based on a Gaussian process emulator of solving the BUU equation to extract $m_{v,0}^*$，using the \textit{surmise} package~\cite{surmise2021,ZZCPC45,Hea2021,LYYNST33,PLGNST34,PDJPG49}.
Specifically, we first calculate $\Gamma$ and $m_1$ using $20$ Skyrme interactions (i.e., BSk1, BSk4, BSk14, BSk15, BSk17, LNS5, MSL1, SIII, SIII*, Skyrme 1', SLy0, SLy1, SLy2, SLy4, SLy5, SLy7, SLyIII.0.7, SQMC650, SV-mas07, SV-sym34)~\cite{DutPRC85, GamPRC84, ZZPLB726, PetNPA584, WasPRC86}, with various $k$ values ranging from $0.0$ to $1.0$ in steps of $0.1$.
The values of $m^{*}_{v,0}/m$ of the above interactions range from $0.51$ to $1.05$.
The above results on $\Gamma$ and $m_1$ provide a representative sampling of the BUU calculation in the parameter space spanned by $m^{*}_{v,0}/m$ and $k$.
They are then used to train a Gaussian process emulator, which enables efficient Markov Chain Monte Carlo~(MCMC) sampling in the $m^{*}_{v,0}/m$--$k$ space.
In the Bayesian analysis, we assume uniform priors for $m^{*}_{v,0}/m$ and $k$ ranging from $0.6$--$0.9$ and $0.1$--$0.6$, respectively. 
The likelihood function is assumed to be the commonly used Gaussian form constructed using the experimental values of $\Gamma = 4.08 \pm 0.09$ MeV~\cite{PluADNDT123} and $m_1 = 883.05\pm 27.03\ \rm{MeV\ fm}^2$~\cite{DieADNDT38,RocPRC92,TamPRL107}. 
A total of $5\times 10^6$ samples in the $m_{v,0}^*/m$--$k$ space are generated from MCMC sampling. 
The posterior distributions of $m^{*}_{v,0}/m$ and $k$ are then estimated using these samples which are shown in Fig.~\ref{F:Q4}. 
The bivariate scatter histograms of the MCMC samples are exhibited in Fig.~\ref{F:Q4}~(b) along with the $68.3\%$ and $90\%$ confidence regions enclosed by the solid lines. 
The corresponding marginal posterior distributions of $m^{*}_{v,0}/m$ and $k$ are displayed in Fig.\ref{F:Q4}~(a) and Fig.\ref{F:Q4}~(c), respectively.

Quantitatively, at $68.3\%$ confidence level we infer
\begin{equation}\label{E:bayes}
\begin{split}
&k = 0.341_{-0.074}^{+0.077},\\
&m^{*}_{v,0}/m = 0.731^{+0.027}_{-0.023}.
\end{split}  
\end{equation}
Combining the inferred $m_{v,0}^*/m$ and the constraint on the isocalar effective mass $m^{*}_{s,0}/m$ $=$ 0.82 $\pm 0.03$ from the BUU analysis of ISGQR of $\isotope[208]{Pb}$ reported in Ref.~\cite{SYDPRC104}.
The linear neutron-proton effective mass splitting coefficient $\Delta m^{*}_{1}(\rho_0)$ is derived to be,
\begin{equation}\label{E:dm1}
\Delta{m^{*}_{1}}(\rho_0)/m = 0.200^{+0.101}_{-0.094}.
\end{equation}
The uncertainties of $\Delta m^{*}_{1}(\rho_0)$ are propagated from the uncertainties of $m^{*}_{v,0}/m$ and $m^{*}_{s,0}/m$.

\begin{figure}[htbp]
%\centering
\includegraphics%[width=\columnwidth]
[width=\linewidth]{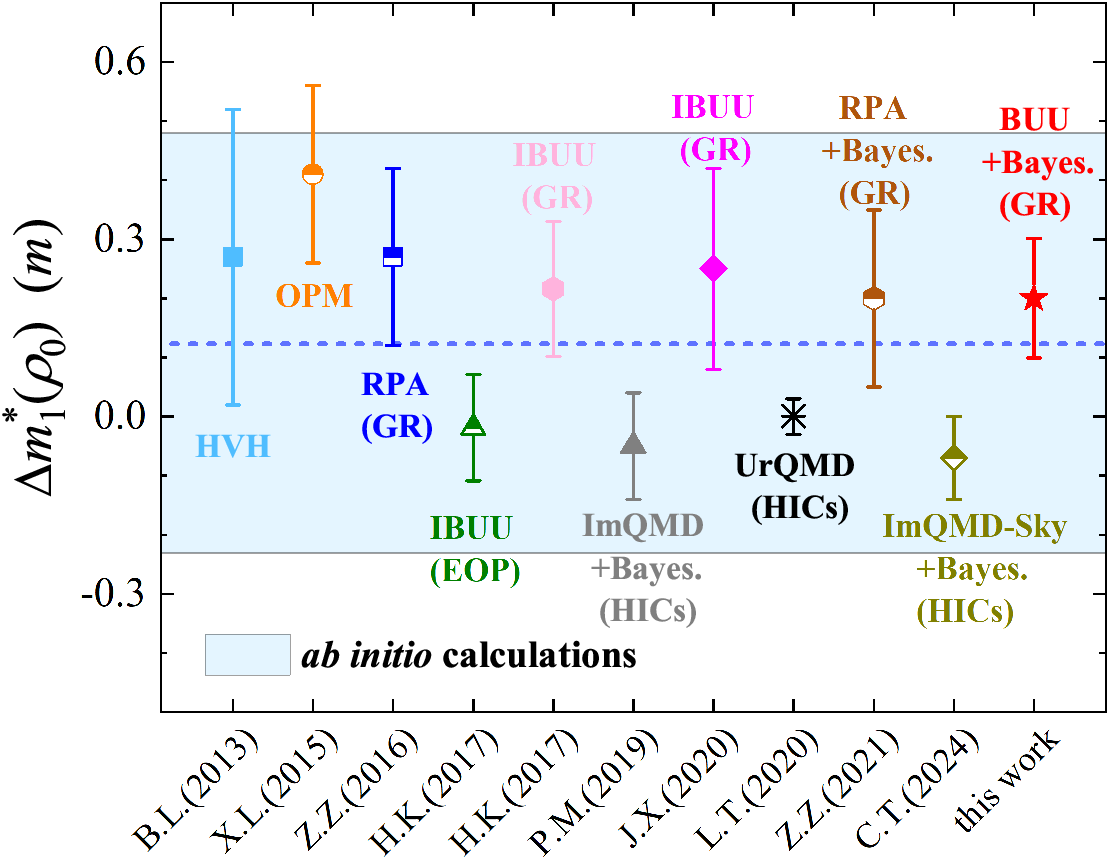}
\caption{
The $\Delta m^{*}_{1}(\rho_0)$ obtained using the BUU equation~(red solid star), is compared with constraints derived from the nuclear structure and nuclear reaction experiment observables~(other symbols, see text for details)~\cite{LBAPLB727, LXHPLB743, ZZPRC93, KHYPRC95, MorPLB799, XJPRC102, TLCPC44, ZZCPC45, TsaPLB853}, as well as theoretical results from several $ab$ $initio$ calculations~(blue shaded regions)~\cite{WSBPRC108}.
} \label{F:Q5}
\end{figure}

Finally, in Fig.~\ref{F:Q5}, we compare the $\Delta m^{*}_{1}(\rho_0)$ obtained in the present work (BUU+Bayes., red solid star) with constraints from other experimental analyses and theoretical calculations.
The solid light blue square in Fig.~\ref{F:Q5} represents the result based on the Hugenholtz–Van Hove theorem~(HVH), derived from the systematics of nuclear symmetry energy and its slope~\cite{LBAPLB727}.
The half-filled yellow circle denotes the constraint obtained from the optical potential model~(OPM) analysis based on nucleon-nucleus scattering data~\cite{LXHPLB743}.
The half-filled blue square, solid pink hexagon, solid peach diamond, and half-filled brown hexagon represent results extracted from experimental observables of nuclear giant resonance~(GR) using the RPA theory~\cite{ZZPRC93}, the isospin-dependent BoltzmannUehling-Uhlenbeck transport model~(IBUU)~\cite{KHYPRC95, XJPRC102}, and the Bayesian analysis based on RPA theory~(RPA+Bayes.)~\cite{ZZCPC45}, respectively.
The half-filled green triangle represents constraints from IBUU model analyses of the empirical optical potential~(EOP)~\cite{KHYPRC95}.
Additionally, the solid gray triangle, black asterisk, and half-filled fulvous diamond, are constraints extracted from experimental observables of HICs through the Bayesian analysis based on the improved quantum molecular dynamics model~(ImQMD+Bayes.)~\cite{MorPLB799}, the ultra-relativistic quantum molecular dynamics model~(UrQMD)~\cite{TLCPC44}, and the Bayesian analysis based on the improved quantum molecular dynamics-Skyrme transport model~(ImQMD-Sky+Bayes.)~\cite{TsaPLB853}, respectively.
Notably, all the three analyses favor negative $\Delta m^{*}_{1}(\rho_0)$.
All these constraints on $\Delta m^{*}_{1}(\rho_0)$ range from -0.13$m$ to 0.56$m$, exhibiting a certain model dependence. 
Despite the values extracted from experimental observations exhibit quantitative---even qualitative---uncertainties, all results fall within the predicted range of various nuclear $ab$ $initio$ many-body calculations, shown as blue shaded regions in Fig.~\ref{F:Q5}.
More details on these $ab$ $initio$ calculations can be found in Ref.~\cite{BalPRC89,WSBPRC108}.
The average of all the presented experimental constraints is $0.12m$, which is indicated by the horizontal dashed line in Fig.~\ref{F:Q5}.
We would like to mention that over the past few decades, significant efforts, both in theory and experiment, have been devoted to determining the neutron-proton effective mass splitting, and those shown in Fig.~\ref{F:Q5} represent only a subset of the related results.

It is worth noting that the analyses based on nuclear collective motions tend to favor a positive $\Delta m^{*}_{1}(\rho_0)$, while those based on HIC observables suggest negative values.
This discrepancy appears to imply a momentum-dependent effective mass splitting, i.e., for low-momentum nucleons---which dominate the dynamics in nuclear collective motions and in HICs with lower incident energies---one has $m^*_n$ $>$ $m^*_p$, whereas for high-momentum nucleons---which play a more significant role in HICs at higher incident energies---the relation reverses, with $m^*_n$ $<$ $m^*_p$.
Due to the limitation of the standard Skyrme EDF, the effective mass obtained with Skyrme interactions shows no momentum dependence.
The momentum-independent $\Delta m^{*}_{1}(\rho_0)$ can be regarded as an overall estimate of the momentum-dependent effective mass splitting averaged over the dominant momentum region relevant to a given process.
A more sophisticated constraint should be imposed directly on the momentum dependence of the effective mass splitting, or equivalently, on the first-order symmetry potential defined as $U_{{\rm sym},1}(p,\rho)$ $\equiv$ $\frac{\partial\epsilon_n(p,\rho,\delta)}{\partial\delta}\Big|_{\delta=0}$ $=$ $-\frac{\partial\epsilon_p(p,\rho,\delta)}{\partial\delta}\Big|_{\delta=0}$.
This could be explored by employing, for example, the Skyrme interactions extended to include higher-order derivative terms (i.e., relative-momentum terms)~\cite{CarPRC78,RaiPRC83,WRPRC98,WSPPRC109,YJPRC109,WSPPRC111,YJTApJ985}.

\section{Summary}\label{4}

In the present work, we employ the BUU equation to obtain the strength function $S(E)$ of IVGDR of $\isotope[208]{Pb}$. 
The careful numerical treatments of the adoption of a large number of ensembles when solving the BUU equation enable an accurate description of the properties of nuclear collective motions based on the semi-classical approximation to the Wigner representation of the time-dependent quantum approach.
The sensitivities of the IVGDR observables, i.e., the full width at half maximum $\Gamma$ of its strength function $S(E)$ and the energy-weighted sum rule $m_1$, to isovector effective mass $m^*_{v,0}/m$ and in-medium NN cross section $\sigma^*$ are investigated.
Based on the Bayesian inference using the BUU-calculated values of $m_1$ and $\Gamma$ as training data, we constrain constant in-medium correction factor $k$ $=$ $0.341^{+0.077}_{-0.074}$ and $m^{*}_{v,0}/m$ $=$ $0.731^{+0.028}_{-0.023}$.
By incorporating a previously constrained value of the isoscalar effective mass $m^{*}_{s,0}/m$ $=$ 0.820 $\pm$ 0.030, we deduce the linear neutron-proton effective mass splitting coefficient $\Delta m^{*}_{1}(\rho_0)/m$ $=$ $0.200^{+0.101}_{-0.094}$.
This result is compared with constraints obtained from other phenomenological approaches and $ab$ $initio$ calculations.
The constraint obtained from the BUU equation shows good agreement with values extracted from experimental observations, which might help understand the underlying  characteristics of $m^{*}_{n-p}$ in neutron-rich nuclear matter, particularly its behavior for low-momentum nucleons.

\begin{acknowledgments}
We thank Chen Zhong for setting up and maintaining the GPU server.
This work is partially supported by the National Natural Science Foundation of China under Grant Nos. $12405146$,   $12235010$, and $12147101$, the Guangdong Major Project of Basic and Applied Basic Research No. 2020B0301030008, and the STCSM under Grant No. 23590780100.
Rui Wang and Zhen Zhang acknowledge the hospitality and support of the Shanghai Research Center for Theoretical Nuclear Physics, where part of this work was discussed.
 
%and the Key Research Program of the CAS under Grant No. XDB34000000.
%No. $12405146$ is the National Natural Science Foundation Youth Fund of SYD.
%No. $12235010$ is the National Natural Science Foundation of ZZ.
%No. $12147101$ is the Shanghai Research Center for Theoretical Nuclear Physics. 

\end{acknowledgments}
\end{CJK*}
\bibliography{Ref}

\end{document}